\documentclass[twocolumn,showpacs,preprintnumbers,amsmath,amssymb,natbib,noshowpacs]{revtex4-2}

\usepackage{amssymb}
\usepackage{graphicx}
\usepackage{tikz}

\usepackage{bm}
\usepackage{amsmath,amsfonts,bm,dsfont}
\usepackage{mathrsfs}         
\makeatletter
\def\l@subsection#1#2{}
\def\l@subsubsection#1#2{}
\makeatother

\usepackage{color}

\newcommand{\red}[1]{{#1}}

\newcommand{\bs}{\boldsymbol}

\newcommand{\add}[1]{{#1}}
\newcommand{\vv}[1]{{#1}}

\newcommand{\be}{\begin{eqnarray}}
	\newcommand{\ee}{\end{eqnarray}}

\newcommand{\rev}[1]{{#1}}
\newcommand{\rv}[1]{{#1}}
\newcommand{\rvv}[1]{{#1}}
\newcommand{\rr}[1]{{#1}}
\newcommand{\zzz}[1]{{#1}}
\newcommand{\zz}[1]{{#1}}

\begin{document}

	\title{Anomalous fractional quantum Hall effect and multi-valued Hamiltonians}
	
	\author{Xi Wu}
	
	\affiliation{Physics Department, Ariel University, Ariel 40700, Israel}
	
	\author{M.A. Zubkov }
	
	\affiliation{Physics Department, Ariel University, Ariel 40700, Israel}
	\affiliation{{\it On leave of absence from} NRC "Kurchatov Institute" - ITEP, B. Cheremushkinskaya 25, Moscow, 117259, Russia}
	
	\vspace{10pt}

	\begin{abstract}
		We discuss anomalous fractional quantum Hall effect that exists without external magnetic field. We propose that excitations in such systems may be described effectively by non-interacting particles with the Hamiltonians defined on the Brillouin zone with a branch cut. Hall conductivity of such a system is expressed through the one-particle Green function. We demonstrate that for the Hamiltonians of the proposed type this expression takes fractional values times Klitzing constant. Possible relation of the proposed construction with degeneracy of ground state is discussed as well.
	\end{abstract}
	
	%
	%
	%
	%
	%

\maketitle

\section{Introduction}

According to the common lore the quantum Hall effect~\cite{Klitzing} \cite{Tsui} has topological reason. Relation to topology has been well established for the case of integer quantum Hall effect (IQHE). 
It is typically associated with the so-called TKNN (Thouless,  Kohmoto,  Nightingale, den Nijs) invariant \cite{TKNN,Fradkin,Tong:2016kpv,Hatsugai,Hall3DTI}.  The corresponding expression is not changed when the system is modified smoothly. This relation, has been obtained originally for the idealized systems in the presence of constant magnetic fields, in the absence of disorder, and without inter-electron interactions. The influence of both disorder and interactions on the Hall conductivity in external magnetic field has been later discussed widely  \cite{Hall000,TKNN2,Altshuler0,Altshuler}. This consideration has also been limited by the case of the constant magnetic field.

The appearance of IQHE without any magnetic field has been proposed \cite{Haldane} in the framework of the Haldane model and its cousins. In such systems energy bands carry nontrivial Chern numbers~\cite{TKNN}. This type of the quantum Hall effect is called now the intrinsic anomalous quantum Hall effect (AQHE), while the corresponding systems are called Chern insulators.  In the absence of the inter-electron interactions the TKNN invariant for the intrinsic QHE (existing without external magnetic field) was expressed through the momentum space Green's function \cite{Matsuyama:1986us,Volovik0} (see also Chapter 21.2.1 in \cite{Volovik2003}). Moreover, it has been shown that the intrinsic anomalous quantum Hall effect (AQHE) conductivity is given by the expression of    \cite{Matsuyama:1986us,Volovik0}, in which the non-interacting two-point Green function has been substituted by the two point Green function with the interaction corrections \cite{ZZ2019}.

Recently, it has been proposed, that the strongly correlated fermionic systems may possess the FQHE without external magnetic field. This is the  anomalous fractional Hall effect (AFQHE). The corresponding systems are typically called fractional Chern insulators (FCI)  ~\cite{Tang,checkerboard,Neupert,Sheng1,YFWang1,XLQi1,Regnault2,FCI_reviews,FCI_reviews1}. It has been conjectured that in those systems the topological flat bands \cite{Tang,checkerboard,Neupert} may play the role of Landau levels. According to the standard hypothesis interactions result in the fractional filling of those topological bands. Numerical results on the dynamics of the corresponding models have been reported, which propose the existence of AFQHE~\cite{Sheng1,YFWang1,Regnault2,FCI_reviews,FCI_reviews1}.
Discussion of related phenomena may also be found in ~\cite{Xiao,Ghaemi,Lukin,Lukin1,Yannopapas,FLiu,Cooper,Cirac-TFB,Cirac-TFB1,Kapit,NYYao3}.

In the past certain efforts have been invested to find suitable many-body wave functions corresponding to the electron states in the Chern insulators. Those constructions are based on an analogy to the Laughlin trial wave function of the two-dimensional (2D) electron gas in a strong magnetic field~\cite{Laughlin} (see also \cite{Haldane1,Haldane2}), and on its decomposition using Jack polynomial ~\cite{Jacks,Jacks1,GPP1,GPP2,GPP3,GPP4}.
On the language of the trial many-body wave functions a mapping has been proposed between the systems with IQHE and AQHE  \cite{XLQi1,XLQi2,XLQi3,XLQi4,XLQi5}. Extension of this approach to the FQHE/AFQHE correspondence  has been discussed as well \cite{AFQHE_w}.

In the present paper we conjecture the alternative approach to the understanding of the AFQHE in fractional Chern insulators. Instead of finding the suitable many-body trial wave function  we base our consideration on an analogy to the topological invariant responsible for the AQHE in topological insulators composed of the two-point Green function \cite{Volovik0,Volovik2003}. We propose that the electron dynamics in FCI can be described effectively by the noninteracting fermionic excitations carrying electron charge, but having the Hamiltonians of unusual type defined on the Brillouin zone with the branch cut. \zz{As we will see, it is exactly the multivalued Hamiltonian having branch cut that gives Hall conductivity equal to a fractional value times Klitzing constant.} \red{It is well - known that fractional quantum Hall effect in the presence of magnetic field is due to interactions. It is also widely believed that intrinsic fractional Hall effect in fractional Chern insulators is due to interactions. Actually, we do not discuss in our paper the dynamical origin of intrinsic fractional Hall effect. Instead we propose the phenomenological model that describes behavior of fermionic quasiparticles. Those quasiparticles may appear as excitations in strongly correlated systems, and the microscopic reasons for their appearance as well as for appearance of the considered effective multi-valued Hamiltonians remains out of the scope of our paper.}

The paper is organized as follows. In Sect. \ref{SectLat} we recall formulation of the lattice models in momentum space and derivation of the topological representation for Hall conductivity through the one-particle Green function.
{In Sect. \ref{SectWind} we introduce the notion of fractional \zz{winding number} of a curve around a point in a plane.  In Sect. \ref{MHC} we \zz{extend consideration of Sect. \ref{SectWind} to}  the simplest example of the Hamiltonian defined on the Brillouin zone with the branch cut. \zz{We show that in this model Hall conductivity is proportional to fractional number}. In Sect. \ref{SectWL} we \red{present the main result of our paper. Namely, we } propose a way to write down Hamiltonians with the branch cut in Brillouin zone that give \zz{the expected fractional quantum Hall effect}. \rr{In Sect. \ref{SectWH} we give} two explicit examples based on Wilson fermion Hamiltonian and Haldane model of topological insulator. In Sect. \ref{SectConcl} we end with the conclusions.}

\section{Hall conductivity as the topological invariant in momentum space}

\label{SectLat}

In this section for completeness we remind briefly the derivation of the topological expression for the conductivity of intrinsic anomalous quantum Hall effect in two-dimensional topological insulators. The Hall conductivity is expressed here through the two-point Green function. Such an expression has been proposed in \cite{Matsuyama:1986us,Volovik0}. In the next sections this expression will be used for the description of the effective models of fractional Chern insulators.

Let us start from brief consideration of lattice models in momentum space following the methodology of \cite{Z2016_1,KZ2018,ZK2017} (see also \cite{FZ2019,FSWZZ2019,FZ2019_2} and references therein). This is the extension to the field theory of the quantum-mechanical Wigner-Weyl calculus \cite{1,2,3,4}.
In the absence of the external gauge field  the partition function of the theory defined on the infinite lattice is
\begin{equation}
Z = \int D\bar{\psi}D\psi \, {\rm exp}\Big( - \int_{\cal M} \frac{d^D {p}}{|{\cal M}|} \bar{\psi}^T({p}){\cal G}^{-1}({ p})\psi({p}) \Big)\label{Z1}
\end{equation}
 Here $|{\cal M}|$ is the volume of momentum space $\cal M$, $D$ is the dimensionality of space-time. Without loss of generality we assume here that (imaginary) time is discretized. This results in the finite value of $|{\cal M}|$. At any step of calculations the discretization of time may be taken off.   $\bar{\psi}$ and $\psi$ are the Grassmann-valued fields defined in momentum space $\cal M$. The Green function $\cal G$ is specific for the given system.
In the presence of the {static} external gauge field corresponding to the potential $A(x)$ (up to the terms irrelevant in the low energy effective theory) we may represent the partition function as follows \cite{Z2016_1}
\begin{eqnarray}
Z = \int D\bar{\psi}D\psi \, {\rm exp}\Big( -  \int_{\cal M} \frac{d^D {p}}{|{\cal M}|} \bar{\psi}^T({p})\hat{\cal Q}(i{\partial}_{p},{p})\psi({p}) \Big)\label{Z4}
\end{eqnarray}
Here
\begin{equation}
\hat{\cal Q} = {\cal G}^{-1}({p} - {A}(i{\partial}^{}_{p}))\label{calQM}
\end{equation}
while the pseudo-differential operator ${A}(i\partial_{p})$ is defined as follows. First, we represent the original gauge field ${A}({r})$ as a series in powers of coordinates ${r}$. Next, variable ${r}$ is substituted in this expansion by the operator $i\partial_{p}$. Besides, in Eq. (\ref{calQM}) each product of the components of ${p} - {A}(i{\partial}^{}_{p})$ is subsitituted by the symmetric combination (for the details see \cite{Z2016_1}).


We relate operator $\hat{Q} = Q(p-A(i\partial_p))$ and its inverse $\hat{G} = \hat{Q}^{-1}$ defined in Hilbert space ${\cal H}$ of functions (on $\cal M$) with their matrix elements ${\cal Q}(p,q)$ and ${\cal G}(p,q)$ correspondingly:
$$
{\cal Q}(p,q) = \langle p|\hat{Q}| q\rangle, \quad {\cal G}(p,q) = \langle p|\hat{Q}^{-1}| q\rangle\,.
$$
Here the basis elements of $\cal H$ are normalized as $\langle p| q\rangle = \delta^{(D)}(p-q)$. Those operators obey the following equation
$$
\langle p|\hat{Q}\hat{G}|q\rangle = \delta({p} - {q})\,.
$$

The Green function of Bloch electron is given by
\rev{\begin{align}
{\cal G}_{ab}(k_1,k_2)&= -\frac{1}{Z}\int D\bar{\psi}D\psi \, {\rm exp}\Big(  -  \int_{\cal M} \frac{d^D {p}}{|{\cal M}|} \bar{\psi}^T({p})\hat{\cal Q}(i{\partial}_{p},{p})\psi({p}) \Big)\nonumber\\
& \frac{\bar{\psi}_b(k_2)}{\sqrt{|{\cal M}|}} \frac{\psi_a(k_1)}{\sqrt{|{\cal M}|}}\label{G1}\,.
\end{align}}
Here indices $a,b$ enumerate the components of the fermionic fields. In the following we will omit those indices for brevity.
\rvv{Notice, that we use the relativistic units, in which both $\hbar $ and $c$ are equal to unity.  Besides, elementary charge $e$ is included to the definition of electric and magnetic fields. }
The Wigner transformation of $\cal G$ is defined as the Weyl symbol of $\hat{G}$:
\begin{equation} \begin{aligned}
{G}_W(x,p) \equiv \int_{\cal M} dq e^{ix q} {\cal G}({p+q/2}, {p-q/2})\label{GWx}
\end{aligned}\,.
\end{equation}
\rvv{Correspondingly, the Weyl symbol of operator $\hat{Q}$ is given by ${Q}_W(x,p) \equiv \int_{\cal M} dq e^{ix q} {\cal Q}({p+q/2}, {p-q/2})$. It appears that for the slowly varying field $A(x)$ we have ${Q}_W(x,p) = {Q}_W(p-A(x))\equiv {Q}(p-A(x))$ (see \cite{Z2016_1}).}
\rv{It is assumed here that $Q(p_1,p_2)$ is nonzero for the values of $|p_1-p_2|$ much smaller than the size of the Brillouin zone. (The values of $|p_1 + p_2|$ may be arbitrary.) This occurs if the external electromagnetic field is slowly varying, i.e. its variation on the distance of the order of the interatomic distance may be neglected. Under these conditions the Wigner transformed Green function obeys the Groenewold equation (see \cite{Z2016_1})}:
\begin{equation}\begin{aligned}
&{G}_W(x_n,p)
e^{\frac{i}{2} \left( \overleftarrow{\partial}_{x_n}\overrightarrow{\partial_p}-\overleftarrow{\partial_p}\overrightarrow{\partial}_{x_n}\right )}
Q_W(x_n,p)= 1
\label{GQW}\end{aligned}\,.
\end{equation}
By $x_n$ we denote the lattice points. Although the lattice points are discrete, the differentiation over $x_n$ may be defined following \cite{Z2016_1} because the functions of coordinates may be extended to their continuous values.

Variation of partition function gives the following expression for the electric current density:
\rev{\begin{eqnarray}
\langle j^k \rangle &=& -  \, \int \frac{d^D p}{(2\pi)^D}\, {\rm Tr} \, G_W(p,x)  \partial_{p_k} Q_W(p - { A}(x)). \label{J3}
\end{eqnarray}}


\label{SectWigner}

Let us consider the case of constant external field strength $A_{ij}$. We are going to expand $\langle j \rangle$ in powers of $A_{ij}$ and to keep the linear term only.
We define $G^{(0)}_W$ that obeys
$$
G^{(0)}_W(p)  Q^{(0)}_W(p)=1\,.
$$
Here $Q^{(0)}_W(p) = {\cal G}^{-1}(p)$, which gives $G^{(0)}_W(p)= {\cal G}(p)$. We use both expansion in powers of $A$ and the derivative expansion. Solution of the Groenewold equation up to the terms linear in $A$ and its derivatives is given by
\begin{eqnarray}
 G_W(p,x)  &= & G_W^{(0)}\nonumber\\&& + G^{(0)}_W \partial_{p_m} Q^{(0)}_W   G^{(0)}_W A_m \nonumber\\&&+ \frac{i}{2}  G_W^{(0)} \frac{\partial  Q^{(0)}_W}{\partial p_i}  G_W^{(0)}  \frac{\partial  Q^{(0)}_W}{\partial p_j}  G_W^{(0)}
A_{ij}\,.
\end{eqnarray}
We then rewrite Eq. (\ref{J3}) as follows:
\rev{\begin{eqnarray}
\langle j^k \rangle  &\approx & - \, \int \frac{d^D p}{(2\pi)^D}\, {\rm Tr} \, G^{(0)}_W(p)  \partial_{p_k}  Q^{(0)}_W(p)  \nonumber\\ && -  \frac{i }{2}\,  \int \frac{d^D p}{(2\pi)^D}\, {\rm Tr} \,  G^{(0)}_W \frac{\partial  Q^{(0)}_W}{\partial p_i}\nonumber\\ &&  G^{(0)}_W \frac{\partial  Q^{(0)}_W}{\partial p_j} G^{(0)}_W  \partial_{p_k} Q^{(0)}_W
A_{ij}\,.
\label{J32}
\end{eqnarray}}
\rv{The first term here is the equilibrium ground state current in the absence of electric field.  It is widely believed to be equal to zero due to the extension of the Bloch theorem to the field-theoretical systems. Anyway, this term cannot contribute to the electric current of the QHE because it does not contain external electric field.  }

In order to obtain expression for Hall current we substitute to the above expressions the Euclidean field strength $A_{ij}$ corresponding to the electric field $E_k$. Its nonzero components are \begin{equation}
A_{Dk} = \partial_D A_k-\partial_k A_D = -i E_k.
\end{equation}
This results in the following expression for the current density in the presence of the external electric field $E_i$:
\begin{eqnarray}
\langle j^{k} \rangle &=& - \frac{1}{2\pi} {\cal N} \epsilon^{3kj} E_{j}\label{j02},\,\nonumber\\
{\cal N} &= & \frac{\epsilon_{ijk}}{\,3!\,4\pi^2} \int d^3p      \, {\rm Tr}  \, \Big[  {\cal G} \frac{\partial {\cal G}^{-1}}{\partial p_i} \frac{\partial  {\cal G}}{\partial p_j} \frac{\partial  {\cal G}^{-1}}{\partial p_k} \Big]\,.\label{calN}
\end{eqnarray}
This is the result obtained in \cite{Matsuyama:1986us,Volovik0} using another methods. It can be shown that for the single valued functions $\cal G$ the above expression for ${\cal N}$ gives integer numbers. Below we will consider the case when $\cal G$ is multi-valued function, i.e. it is defined on the Brillouin zone with the branch cut.

\section{Multivalued functions and fractional winding number}

\label{SectWind}

Before we come to the consideration of the models with multi-valued Green function in momentum space let us discuss the appearance of the fractional winding number in a {simpler} case.

We will introduce here the notion of fractional winding number resulted from multi-valued functions as a generalization from the usual integer winding number related to a single-valued function. More explicitly, let us consider a one dimensional closed curve $(x(t), y(t))$ in two-dimensional plane parametrized by parameter $t$. Both functions $x(t)$ and $y(t)$ are periodic with period $T$, so that $t$ is identified with $t+T$:
\be
	 \left\{
  \begin{array}{lr}
    x(t)=x(t+T) &\\
    y(t)=y(t+T) &
  \end{array}
\right..
\ee
The winding number of this curve around the point $(0,0)$ is given by
\be
	N_1:=\frac{1}{2\pi}\int_{t=t_0}^{t=t_0+T}\frac{xdy/dt-ydx/dt}{x^2+y^2}dt\,.
\ee
Another convenient parametrization is
\be
	 \left\{
  \begin{array}{lr}
    z(t)=x(t)+iy(t) &\\
    \bar{z}(t)=x(t)-iy(t) &
  \end{array}
\right. \Leftrightarrow \quad
\left\{
  \begin{array}{lr}
    x(t)=\frac{1}{2}(z(t)+\bar{z}(t)) &\\
    y(t)=\frac{1}{2i}(z(t)-\bar{z}(t)) &
  \end{array}
\right.\,,
\ee
and in this way the winding number can be conveniently rewritten as
\be\label{N1zz}
	N_1&=&\frac{1}{4\pi i}\int_{t=t_0}^{t=t_0+T}\frac{\bar{z}dz/dt-zd\bar{z}/dt}{z\bar{z}}dt\nonumber\\
	&=&\frac{1}{4\pi i}\int_{t=t_0}^{t=t_0+T}d \ln(\frac{z(t)}{\bar{z}(t)})\,.
\ee
{The periodicity is expressed as}
\be\label{zzn1}
	 \left\{
  \begin{array}{lr}
    z(t+T)=z(t)e^{2\pi n_1i} &\\
    \bar{z}(t+T)=\bar{z}(t)e^{-2\pi n_1i} &
  \end{array}
\right.,\,
\text{where }
n_1 \in {Z}\,.
\ee
Substituting {Eq. (\ref{zzn1}) into Eq. (\ref{N1zz})}, we obtain the integer winding number
\be
	N_1=n_1\,.
\ee

\begin{figure}[h!]  \centering     \includegraphics[width=0.7\columnwidth]{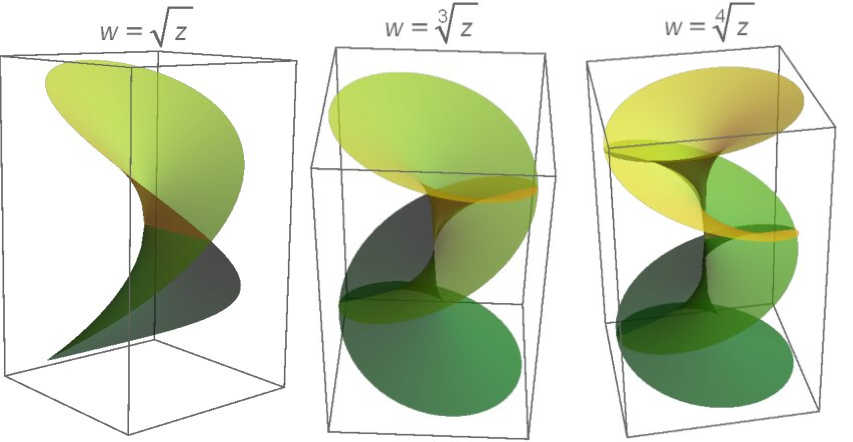}     \caption{ \label{fig000}{We present here the Riemann surfaces of functions $z^{1/n}$ with $n = 2,3,4$. }} \end{figure}

Now let us come to the consideration of multi-valued functions {$z'(t)$ and $\bar{z}'(t)$ composed of the above single-valued functions $z(t)$ and $\bar{z}(t)$:
\be
	 \left\{
  \begin{array}{lr}
    z'(t)=z^{\frac{1}{n}}(t) &\\
    \bar{z}'(t)=\bar{z}^{\frac{1}{n}}(t) &
  \end{array}
\right. \& \quad
\left\{
  \begin{array}{lr}
    x'(t)=\frac{1}{2}(z'(t)+\bar{z}'(t)) &\\
    y'(t)=\frac{1}{2i}(z'(t)-\bar{z}'(t)) &
  \end{array}
\right.\,.
\ee}
Here functions $x'(t)$ and $y'(t)$ define {a} new curve. \rr{Function $z^{1/n}$ is multi-valued. On Fig. \ref{fig000} we represent the corresponding Riemann surfaces for $n=2,3,4$.}
 {Translation} $t \to t + T$ for function $z'(t)$ and $\bar{z}'(t)$ results in
\be
	 \left\{
  \begin{array}{lr}
    z'(t+T)=z'(t)e^{ \frac{2\pi n_1i}{n}} &\\
    \bar{z}'(t+T)=\bar{z}'(t)e^{-\frac{ 2\pi n_1i}{n}} &
  \end{array}
\right.,\,
\ee
with coprime values of $n_1$ and $n$.
One can see the periodicity is lost. Actually, now $z'(t)$ and $\bar{z}'(t)$ are periodic functions with the period $nT$. We can also say that functions $z'(t)$ and $ \bar{z}'(t)$ have $n$ branches in the interval {$t\in (0,nT]$}:
\be
	 \left\{
  \begin{array}{lr}
    t\to z'(t), z'(t)e^{\frac{2\pi i}{n}},..., z'(t)e^{\frac{2\pi (n-1) i}{n}} &\\
 t\to \bar{z}'(t), \bar{z}'(t)e^{-\frac{2\pi i}{n}},..., \bar{z}'(t)e^{\frac{-2\pi (n-1) i}{n}} &
  \end{array}
\right.\,.\label{bc}
\ee
Now we define the notion of fractional winding number for the curve given by functions $x'(t)$ and $y'(t)$.
Each branch is not periodic as a function of $t$ with the period $T$. At $t=t_0$ it has a discontinuity corresponding to the entrance to another branch. We propose the following formal definition of the winding number:
\be\label{N'1}
	N'_1&:=&\frac{1}{4\pi i}\int_{t=t_0}^{t=t_0+T}d \ln(\frac{z'(t)}{\bar{z}'(t)})\nonumber\\
	&=&\frac{1}{n}N_1=\frac{1}{n}\frac{1}{4\pi i}\int_{t=t_0}^{t=t_0+nT}d \ln(\frac{z'(t)}{\bar{z}'(t)})\,.
\ee
As we can see, this definition is independent of the value of $t_0$ which is the branch cut, so it is free from ambiguity.
The meaning of this fractional winding number is that all  $n$ different branches of the curve  $(z'(t),\bar{z}'(t))$ \rr{wind} around the point $(0,0)$ precisely $N_1$ times. Each piece of the curve corresponding to the interval $t\in (t_0, t_0+T]$ effectively winds up $N_1/n$ times. Notice that the generalization to the notion of irrational winding number may be inappropriate since it would require an infinitely large number of branches and thus an infinitely large denominator.

\section{Multi-valued Hamiltonian. The simple example.}
\label{MHC}

 In Sect. \ref{SectLat} we considered the {conventional} models with noninteracting electrons. The obtained expression for the Hall conductivity appears to be the integer multiple of $1/(2\pi)$. The corresponding integer number is the topological number that is a generalization of the winding number. Let us suppose that the system may be expressed qualitatively as the collection of noninteracting excitations. For those excitations we expect the appearance of the Hermitian hamiltonian. Moreover, this hamiltonian is assumed to give an unambiguous expression for the dependence of energy on momentum. But we allow the Hamiltonian itself to be {one branch of} the multivalued function on the Brillouin zone, so that the Brillouin zone contains the {branch} cut. Below the toy models of this type are considered, and it is shown that the corresponding expressions for the Hall conductivity are indeed given by noninteger multiples of $1/(2\pi)$.


In order to better understand relation between the multi-valued functions and non-integer winding numbers we considered in the previous section \ref{SectWind} an example that was not related directly to Hall conductivity. Here the more complicated example is considered related directly to the topological invariant responsible for the FQHE.
We consider the two-dimensional systems with
\be
{\cal G}^{-1}=i\omega-H=i\omega-\sum_{a=1}^{3}h_a\sigma^a\,\ee
Here $H$ is the Hamiltonian while $\omega$ is the Matsubara frequency.
We also denote
\be
{\cal G}=\sum_{a=0}^{3}g_a\sigma^a\,,\quad	{\hat{g}_a=\frac{g_a}{\sqrt{-\det \cal G}}}\,.\nonumber
\ee

According to the results of Sect. \ref{SectLat} the Hall conductivity is given by $\sigma_{H} = \frac{1}{2\pi} N_3$. The corresponding \zz{value of $N_3$} in momentum space in $2+1$ D is given by
\be
	N_3&=&\frac{2\pi }{6}\int \frac{d^3 p}{(2\pi)^3} \epsilon_{ijk}\,\text{tr} ({\cal G} \frac{\partial}{\partial p_i} {\cal G}^{-1}{\cal G} \frac{\partial}{\partial p_j} {\cal G}^{-1} {\cal G} \frac{\partial}{\partial p_k} {\cal G}^{-1})\nonumber
	\\
	&=& \frac{2\pi i}{6}\int \frac{d^3 p}{(2\pi)^3} \epsilon_{ijk}\epsilon^{abcd}\hat{g}_a \frac{\partial}{\partial p_i}\hat{g}_b \frac{\partial}{\partial p_j}\hat{g}_c \frac{\partial}{\partial p_k}\hat{g}_d\,,
\ee
Integrating over $p_0(=\omega)$, we get
{\be \label{hhh}
	N_3&=&\frac{1}{4\pi}\int d^2 \boldsymbol{p} \, \epsilon^{abc} \hat{h}_a \partial_{ p_x}\hat{h}_b\partial_{ p_y}\hat{h}_c\nonumber
	\\
	&=& \frac{1}{4\pi}\int d^2 \boldsymbol{p} \, \epsilon^{abc} \frac{h_a}{|h|^3}\partial_{ p_x}h_b\partial_{p_y}h_c
	\,,
\ee}
where $|h|=\sqrt{\sum_{a=1}^{3}h_a^2}$. 
We will use Eq.(\ref{hhh}) for the calculation of topological invariants if the one-particle Hamiltonian is known.

The general theory developed in Sect. \ref{SectLat} was devoted to the noninteracting fermions defined on the lattice. However, the final expression is also valid for the continuum theory. In this section we consider the continuum toy model with the Hamiltonian
\be\label{H13}
	H_{1/3}&=&\left[\begin{array}{cc}m & (p_1-ip_2)^{1/3} \\(p_1+ip_2)^{1/3} & -m\end{array}\right]\nonumber\\
	&=&\left[\begin{array}{cc}h_3 & h_1-ih_2 \\h_1+ih_2 & -h_3\end{array}\right]\,,
\ee
in which
\begin{align}\label{h|h|}
	|h|&=\sqrt{(p_1^2+p_2^2)^{1/3}+m^2}\,.
\end{align}
{Though not  written explicitly, we only choose one value of $(p_1-ip_2)^{1/3}$ in the definition of this Hamiltonian. Hermiticity requires that the branch of  $(p_1+ip_2)^{1/3}$ is given such that it is complex conjugate of $(p_1-ip_2)^{1/3}$ .}
{The Green function is
\begin{eqnarray}
&&	{\cal G}_{1/3}=(i\omega-H_{1/3})^{-1}\nonumber\\&&=\frac{-1}{\omega^2+(p_1^2+p_2^2)^{1/3}+m^2}\left[\begin{array}{cc}i\omega -m & (p_1-ip_2)^{1/3} \\(p_1+ip_2)^{1/3} & i\omega +m\end{array}\right]\,.\nonumber
\end{eqnarray}
}
The branch points of $h_1 \pm ih_2$ are at $\sqrt{(p_1^2+p_2^2)}=0$ and $\sqrt{(p_1^2+p_2^2)}=\infty$. For the integration, we can choose the branch cut at $\arctan \frac{p_2}{p_1}=0$.
Substituting Eq.(\ref{H13}) into Eq.(\ref{hhh}), we get
\be
	N_3[H_{1/3}]=\frac{1}{4\pi}\int\frac{m}{|h|^3}dh_1(p)\wedge dh_2(p) .
\ee
\rr{Further it becomes
\be
	N_3[H_{1/3}]=\frac{1}{4\pi}\int \frac{dp_1 \wedge dp_2}{9}\frac{m \Big(p_1^2+p_2^2\Big)^{-\frac{2}{3}}}{\Big(\sqrt{(p_1^2+p_2^2)^{1/3}+m^2}\Big)^3}\,.
\ee}
And we deduce
\be
	N_3[H_{1/3}]=\frac{1}{6}\frac{m}{|m|}\,.
\ee
Though this Hamiltonian is a three-fold function in momentum space, it is Hermitian and the energies are single-valued, as promised. \zzz{It is worth mentioning that the value of $\cal N$ for the single continuum massive $2+1$D Dirac fermion with single - valued Hamiltonian is half - integer rather than integer. This is because the continuum Dirac fermions are marginal (see Sect. 11.4.2 in \cite{Volovik2003}). Actually any lattice model always contains pairs of two - component fermions, and therefore the value of $\cal N$ is integer for the single - valued lattice Hamiltonians as we will see in the next sections.}

\section{Multi-valued Hamiltonian. A more general situation.}

\label{SectWL}

From the above sections we see that the fractional winding number comes from the multi-valued functions. The example of Sec. (\ref{MHC}) can be generalized straightforwardly to the case of the lattice model. {For a lattice Hamiltonian $H_1$ \rr{given by} 2-by-2 matrix with integer winding number $N_3[H_1]$, there is always a class of Hamiltonians $H_{k/n}$ with winding number $\frac{k}{n}$ times $N_3[H_1]$ (k and n are {coprime} integers). In this section we prove this relation, namely
\be \label{NknN1}
	N_3[H_{k/n}]=\frac{k}{n} N_3[H_1]\,.
\ee}

Let
\be \label{HknH1}
	H_{k/n}&=&\left[\begin{array}{cc}h_3 & h_1-ih_2 \\ h_1+ih_2 & -h_3\end{array}\right]
	\\
	&:=&\left[\begin{array}{cc}h'_3 & (h'_1-ih'_2)^{\frac{k}{n}} \\ (h'_1+ih'_2)^{\frac{k}{n}} & -h'_3\end{array}\right]\nonumber
	\\
	H_{1}&=&\left[\begin{array}{cc}h'_3 & h'_1-ih'_2 \\ h'_1+ih'_2 & -h'_3\end{array}\right]\,.
\ee
Each of the two off-diagonal components of $H_{k/n}$ have $n$ values, or more accurately $n$ arguments. In mathematical language, {$h_1\pm ih_2$ are in Riemann surfaces of $h'_1\pm ih'_2$} , each having $n$ Riemann sheets. However, 
we do not need to specify which Riemann sheet \rr{is chosen}.

\rr{We have}
\be
	N_3[H_1]&=& \frac{1}{4\pi}\int d^2 \boldsymbol{p} \, \epsilon^{abc} \frac{h'_a}{|h'|^3}\partial_{ p_x}h'_b\partial_{p_y}h'_c\nonumber \\
	&=& \frac{1}{4\pi}\int \frac{1}{|h'(p)|^3}(h'_1(p)dh'_2(p)\wedge dh'_3(p)\nonumber \\
	&&+h'_2(p)dh'_3(p)\wedge dh'_1(p)+ h'_3(p)dh'_1(p)\wedge dh'_2(p))\nonumber \\
	N_3[H_{k/n}]&=& \frac{1}{4\pi}\int d^2 \boldsymbol{p} \, \epsilon^{abc} \frac{h_a}{|h|^3}\partial_{ p_x}h_b\partial_{p_y}h_c \nonumber \\
	&=&\frac{1}{4\pi}\int\frac{1}{|h(p)|^3}(h_1(p)dh_2(p)\wedge dh_3(p)\nonumber \\
	&&+h_2(p) dh_3(p)\wedge dh_1(p)
\nonumber \\
	&&+ h_3(p)dh_1(p)\wedge dh_2(p)) ,
\ee
Let us first prove a useful formula \vv{(see also \cite{Z2016_1,Sticlet:2012aa})
\be\label{N3H1sgn}	
	N_3[H_{1}]&=&\frac{1}{2}\sum_{p_a}\text{sgn}(h'_3){\rm Res} \,(\tilde{\boldsymbol{h}}')|_{p_j=p^a_j}\nonumber\\
	&&\text{with } \tilde{\boldsymbol{h}}'(p^a_j)=0~,~\tilde{\boldsymbol{h}}'=(h'_1,h'_2)\,.
\ee
where 
$$
{\rm Res} \,(\tilde{\boldsymbol{h}}')|_{p_j=p^a_j} = \frac{1}{2\pi i} \int_{\partial \mathbf{a}_{\epsilon}} \frac{d(h'_1+i h'_2)}{h'_1+i h'_2}$$
$$ = \frac{1}{2\pi} \int_{\partial \mathbf{a}_{\epsilon}} d \,{\rm Arg}\,(h'_1+i h'_2)
$$
is an integer winding number of mapping ${\rm Arg}\,(h'_1+i h'_2) : \partial \mathbf{a}_{\epsilon} \to U(1)$, 
and 
 $\mathbf{a}_{\epsilon}= \{(p_1, p_2)|(h_1'(p))^2+(h_2'(p))^2\le \epsilon \}$.}
We use an identity which can be checked easily:
\be
	&&\frac{1}{|h'|^3}(h'_1dh'_2\wedge dh'_3+h'_2dh'_3\wedge dh'_1+ h'_3dh'_1\wedge dh'_2) \nonumber
	\\
	&=& \frac{i}{2} d \frac{h'_3}{|h'|}\wedge d \ln (\frac{h'_1+ih'_2}{h'_1-ih'_2}) \,.
\ee
From this, the topological number $N_3[H_1]$ can be written as
\be\label{N3H1a}
	N_3[H_{1}] &=&\frac{i}{8\pi}\int_{BZ}  d \frac{h'_3}{|h'|}\wedge d \ln (\frac{h'_1+ih'_2}{h'_1-ih'_2})\,.
\ee
\zz{Function} $\ln(\frac{h'_1+ih'_2}{h'_1-ih'_2})$ is not defined everywhere in the Brillouin zone. \zz{Therefore,} for the integration of Eq. (\ref{N3H1a}), we need to remove the singular regions. This can be done as \zz{follows}
\be
	N_3[H_{1}] &=&\sum_{\mathbf{a}}\lim_{\epsilon \to 0}\frac{i}{8\pi}\int_{\zz{BZ \setminus \mathbf{a}_{\epsilon}}}  d \frac{h'_3}{|h'|}\wedge d \ln (\frac{h'_1+ih'_2}{h'_1-ih'_2})\nonumber
	\\
	&=&\sum_{\mathbf{a}}\lim_{\epsilon \to 0}\frac{i}{8\pi}\int_{\zz{BZ \setminus\mathbf{a}_{\epsilon}}}  d (\frac{h'_3}{|h'|} d \ln (\frac{h'_1+ih'_2}{h'_1-ih'_2}))\nonumber
	\\
	&=:&\sum_{\mathbf{a}}\int_{\zz{BZ \setminus \{\mathbf{a}\}}}  d A'\,,
\ee
 Via Stokes theorem, the integration becomes a contour integral:
\begin{align}
	N_3[H_{1}] 	&=\zz{-}\sum_{\mathbf{a}}\zz{\lim_{\epsilon \to 0}\oint_{\partial \mathbf{a}_\epsilon}}   A'\nonumber
	\\
	&=-\sum_{\mathbf{a}}\lim_{\epsilon \to 0}\frac{i}{8\pi}\oint_{\partial \mathbf{a}_{\epsilon}}    \frac{h'_3}{|h'|} d \ln (\frac{h'_1+ih'_2}{h'_1-ih'_2})\nonumber
	\\
	&=-\sum_{\mathbf{a}}\lim_{\epsilon \to 0}\frac{i}{8\pi} \frac{h'_3}{\zzz{\sqrt{|(h_3')^2+\epsilon|}}}\oint_{\partial \mathbf{a}_{\epsilon}}    d \ln (\frac{h'_1+ih'_2}{h'_1-ih'_2})
\,,
\end{align}
The contour integral can be evaluated
\vv{\be\label{deltah}
	&&\frac{i}{2} \oint_{\partial \mathbf{a}_{\epsilon}}    d \ln (\frac{h'_1+ih'_2}{h'_1-ih'_2})\nonumber\\
	&=& i \oint_{\partial \mathbf{a}_{\epsilon}}     \frac{d(h'_1+ih'_2)}{h'_1+ih'_2}
\ee
This results in Eq (\ref{N3H1sgn}) (see also \cite{Duan1993,Duan2000}).}

In a similar manner, we write $N_3[H_{k/n}]$ as
\be\label{N3Hkn}
	N_3[H_{k/n}] &=&\frac{i}{8\pi}\int_{BZ}  d \frac{h_3}{|h|}\wedge d \ln (\frac{h_1+ih_2}{h_1-ih_2}) 
\ee
Moreover, because of Eq. (\ref{HknH1}), globally we have
\be
	&&d \ln (\frac{h_1+ih_2}{h_1-ih_2})=\frac{k}{n} d \ln (\frac{h'_1+ih'_2}{h'_1-ih'_2})\,.
\ee
Again, we remove the singularities
\be\label{N3HknM}
	N_3[H_{k/n}] &=&\sum_{\mathbf{a}}\lim_{\epsilon \to 0}\frac{i}{8\pi}\int_{\zz{BZ \setminus\mathbf{a}_{\epsilon}}}  d \frac{h_3}{|h|}\wedge d \ln (\frac{h_1+ih_2}{h_1-ih_2})\nonumber
	\\
	&=&\sum_{\mathbf{a}}\lim_{\epsilon \to 0}\frac{i}{8\pi}\int_{\zz{BZ \setminus\mathbf{a}_{\epsilon}}} \frac{k}{n} d (\frac{h'_3}{|h|} d \ln (\frac{h'_1+ih'_2}{h'_1-ih'_2}))\nonumber
	\\
	&=:&\sum_{\mathbf{a}}\int_{\zz{BZ \setminus \{\mathbf{a}\}}}  d A\,,
\ee
In addition to the factor $k/n$  there is  $|h|$ instead of $|h'|$ in  the denominator. However, note that though $h_1(p)$ and $h_2(p)$ are multi-valued,  differential one-form $A$ is single-valued.
\zz{Therefore} Eq. (\ref{N3HknM}) becomes
\begin{align}
	&N_3[H_{k/n}] 	=\zz{-}\sum_{\mathbf{a}}\oint_{\partial \mathbf{a}}   A\nonumber
	\\
	&=\zzz{-}\sum_{\mathbf{a}}\lim_{\epsilon \to 0}\frac{i}{8\pi} \frac{h'_3}{\zzz{\sqrt{|(h_3')^2+\epsilon^{k/n}|}}}\oint_{\partial \mathbf{a}_{\epsilon}}    \frac{k}{n}d \ln (\frac{h'_1+ih'_2}{h'_1-ih'_2})
\,,
\end{align}
Following the same procedure as \vv{above, 
 we get
\be\label{N3Hknsgn}	
	N_3[H_{k/n}]&=&\frac{k}{2n}\sum_{p^a} \text{sgn}(h'_3){\rm Res} \,(\tilde{\boldsymbol{h}}')|_{p_j=p^a_j}\nonumber\\
	&&\text{with } \tilde{\boldsymbol{h}}'(p^a_j)=0\,.
\ee}
The difference \rr{between} $|h|$ and $|h'|$ does not appear in the final expression since the values are taken \rr{at the} points, where $|h(p^a)|=|h'(p^a)|=|h_3(p^a)|$.
Comparing with Eq. (\ref{N3H1sgn}), we see that Eq. (\ref{NknN1}) is obtained.

\section{Models of Wilson fermions and Haldane model}
\label{SectWH}

\begin{figure}[h!]   \centering     \includegraphics[width=0.6 \columnwidth]{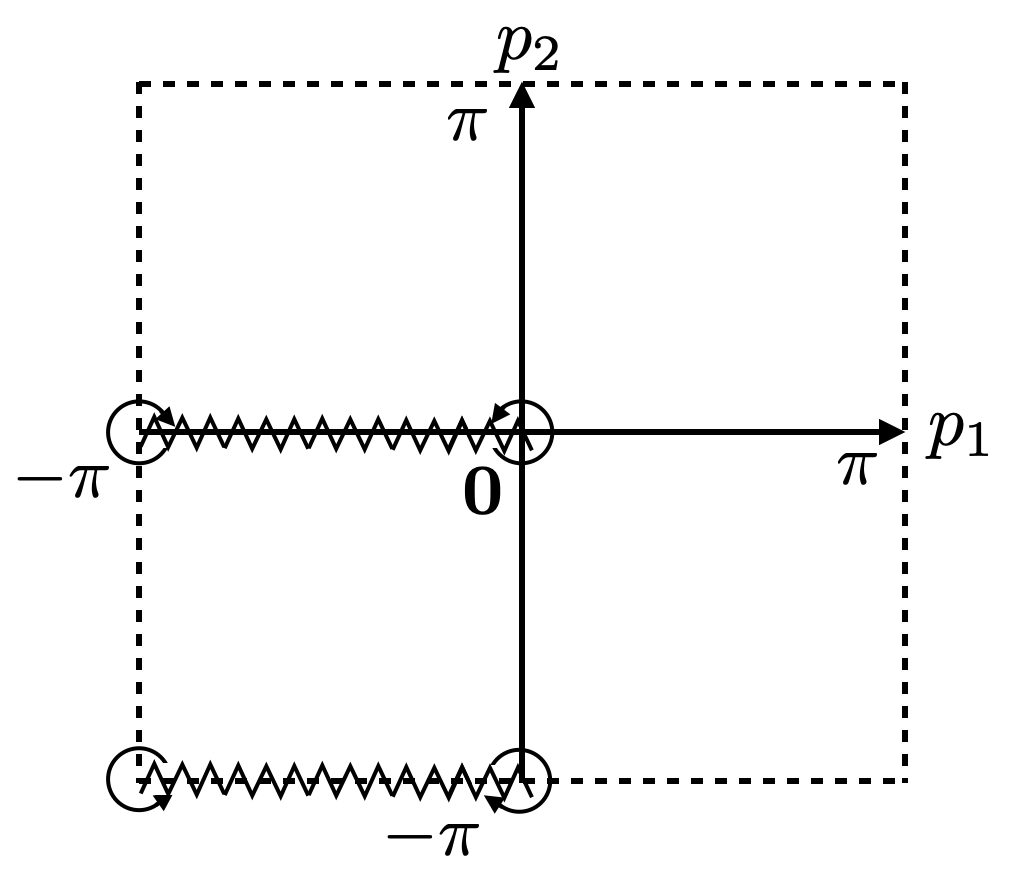}     \caption{ \label{fig001}{In the figure the branch cuts in the Brillouin zone of the model constructed of Wilson fermions are shown. The branch points are obtained from setting $z=\bar{z}=0$. And the branch cuts are obtained by setting $\text{Re}~z<0$, $\text{Im}~z=0$ and $\text{Re}~\bar{z}<0$, $\text{Im}~\bar{z}=0$ , where $z=\sin p_1+i\sin p_2$, $\bar{z}=\sin p_1-i\sin p_2$}.  } \end{figure}

{Let us now apply the general formalism developed above to \rr{certain} particular models. First, let us consider} the Hamiltonian inspired by the model with Wilson fermions. This model is used widely both in lattice quantum field theory understood as the regularization of relativistic quantum field theory and in condensed matter physics. In the latter case it describes qualitatively the topological insulators as well as the Dirac semimetals. The corresponding Hamiltonian has the form
\be
	H_W	&=&\left[\begin{array}{cc}\tilde{m}(p) & \sin p_1-i\sin p_2 \\ \sin p_1+i\sin p_2 & -\tilde{m}(p) \end{array}\right]\,,
\ee
where $\tilde{m}(p)=m+2-\cos p_1-\cos p_2$.
From above we construct
\begin{align}
H_{W,\frac{1}{n}}&=\left[\begin{array}{cc}\tilde{m}(p) & (\sin p_1-i\sin p_2)^{\frac{1}{n}} \\ (\sin p_1+i\sin p_2)^{\frac{1}{n}} & -\tilde{m}(p)\end{array}\right]\,.
\end{align}
{Substituting it into Eq. (\ref{N3Hknsgn}), we get
\begin{align}
	N_3[H_{W,\frac{1}{n}}]=\frac{\text{sgn}(m+4)+\text{sgn}(m)-2\text{sgn}(m+2)}{2n}\,.
\end{align}
For $m<-4$ or $m>0$ we have $N_3[H_{W,\frac{1}{n}}]={0}$; for $-4<m<-2$ we have $N_3[H_{W,\frac{1}{n}}]={1/n}$; for $-2<m<0$ we have $N_3[H_{W,\frac{1}{n}}]={-1/n}$.} \zz{In Fig. \ref{fig001} we represent the brunch cuts in the Brillouin zone of this model.}

{The second model we look at is based on the Haldane model, perhaps the first model realizing anomalous quantum Hall effect. The Hamiltonian reads:
\begin{align}
	&H_H=2t_2\cos\phi\sum_i\cos(\boldsymbol{p}\cdot \boldsymbol{b}_i)\sigma_0+t_1\sum_i[\cos(\boldsymbol{p}\cdot \boldsymbol{a}_i)\sigma_1\nonumber\\
	&+\sin(\boldsymbol{p}\cdot \boldsymbol{a}_i)\sigma_2]+[M-2t_2\sin\phi\sum_i\sin(\boldsymbol{p}\cdot \boldsymbol{b}_i)]\sigma_3\nonumber\\
	&=2t_2\cos\phi\sum_i\cos(\boldsymbol{p}\cdot \boldsymbol{b}_i)\sigma_0\nonumber\\&+[M-2t_2\sin\phi\sum_i\sin(\boldsymbol{p}\cdot \boldsymbol{b}_i)]\sigma_3\nonumber\\
	&+\left[\begin{array}{cc}0 & t_1\sum_i e^{-i\boldsymbol{p}\cdot \boldsymbol{a}_i} \\ t_1\sum_i e^{i\boldsymbol{p}\cdot \boldsymbol{a}_i}& 0\end{array}\right]\nonumber\\
	&=\left[\begin{array}{cc}h'_0+h'_3 & h'_1-ih'_2 \\ h'_1+ih'_2& h'_0-h'_3\end{array}\right]\,,
\end{align}
where
\begin{align}
	&|\boldsymbol{a}_1|=|\boldsymbol{a}_2|=|\boldsymbol{a}_3|=a\nonumber
	\\
	&\cos  \langle \boldsymbol{a}_1,\boldsymbol{a}_2\rangle=\cos \langle \boldsymbol{a}_2,\boldsymbol{a}_3\rangle=\cos \langle \boldsymbol{a}_3,\boldsymbol{a}_1\rangle=-\frac{1}{2}\nonumber
	\\	
	&\boldsymbol{b}_i=\frac{1}{2}\epsilon_{ijk}( \boldsymbol{a}_j-\boldsymbol{a}_k)\,.
\end{align}
\rr{Our fractional topological insulator made of this model is defined by Hamiltonian}
\begin{align}
	H_{H,\frac{1}{n}}&=2t_2\cos\phi\sum_i\cos(\boldsymbol{p}\cdot \boldsymbol{b}_i)\sigma_0\nonumber\\
	&+[M-2t_2\sin\phi\sum_i\sin(\boldsymbol{p}\cdot \boldsymbol{b}_i)]\sigma_3\nonumber\\
	&+\left[\begin{array}{cc}0 & (t_1\sum_i e^{-i\boldsymbol{p}\cdot \boldsymbol{a}_i})^{\frac{1}{n}} \\ (t_1\sum_i e^{i\boldsymbol{p}\cdot \boldsymbol{a}_i})^{\frac{1}{n}} & 0\end{array}\right]\nonumber\\
	&=\left[\begin{array}{cc}h_0+h_3 & h_1-ih_2 \\ h_1+ih_2& h_0-h_3\end{array}\right]\,.
\end{align}
\rr{After some algebra} we get the expression for the determinant
\be
	\det (\frac{\partial \tilde{h}'_i}{\partial p_j})=-\frac{\sqrt{3}}{2}t^2_1a^2\sum_i\sin(\boldsymbol{p}^{\pm}\cdot\boldsymbol{b}_i)\,.
\ee
Next, setting $$h'_1=h'_2=0$$ gives
\be
	\sum_i e^{i\boldsymbol{p}\cdot \boldsymbol{a}_i}=\sum_i e^{-i\boldsymbol{p}\cdot \boldsymbol{a}_i}=0\,,
\ee
which leads to
\be
   &&\sin(\boldsymbol{p}^{\pm}\cdot (\boldsymbol{a}_1-\boldsymbol{a}_2))= \sin(\boldsymbol{p}^{\pm}\cdot (\boldsymbol{a}_2-\boldsymbol{a}_3)) \nonumber
   \\
  && =\sin(\boldsymbol{p}^{\pm}\cdot (\boldsymbol{a}_3-\boldsymbol{a}_1))=\pm\frac{\sqrt{3}}{2}
\ee
thus
\be
	\sin(\boldsymbol{p}^{\pm}\cdot \boldsymbol{b}_i)=\pm\frac{\sqrt{3}}{2}\,.
\ee
The above equation allows us to obtain the result of Eq. (\ref{N3Hknsgn})
\be
	N_3[H_{H,\frac{1}{n}}]=\frac{1}{2n}(\text{sgn}(M+3\sqrt{3}t_2\sin\phi)\nonumber\\
	-(\text{sgn}(M-3\sqrt{3}t_2\sin\phi))\,.
\ee
\zz{In Fig. \ref{fig002} we represent the {branch} cuts in the Brillouin zone of this model.}

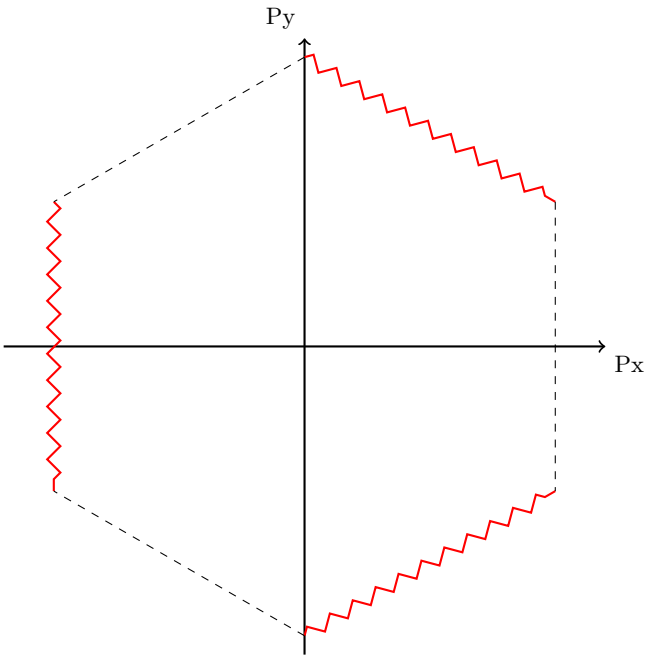
\begin{figure}[h!]   \centering
	
\begin{tikzpicture}\centering
	\usetikzlibrary{snakes}
	\draw[thick,->] (-4,0) -- (4,0) node[anchor=north west] {Px};
	\draw[thick,->] (0,-4.1) -- (0,4.1)node[anchor=south east] {Py};
	\draw[red,thick,snake=zigzag] (0,-3.849) -- (3.33333,-1.9245);
	\draw[red,thick,snake=zigzag] (0,3.849) -- (3.33333,1.9245);
	\draw[red,thick,snake=zigzag] (-3.33333,1.9245) -- (-3.33333,-1.9245);
	\draw[dashed] (0,-3.849) -- (-3.33333,-1.9245);
	\draw[dashed] (0,3.849) -- (-3.33333,1.9245);
	\draw[dashed] (3.33333,1.9245) -- (3.33333,-1.9245);
\end{tikzpicture}
\caption{ In this figure the branch cuts in the Brillouin zone of Haldane model are shown.  We choose the branch cuts as the zigzag lines connecting two types of branch points, which correspond to $z=0$, where $z=\sum_i e^{i\boldsymbol{p}\cdot \boldsymbol{a}_i}$.}
\label{fig002}

\end{figure}

}


\section{Conclusions and discussions}

\label{SectConcl}

In this paper we propose scenario, in which collective excitations in fractional Chern insulators carry electric charge of electron. We suppose, that the dynamics of quantum Hall effect in some of those materials may be described qualitatively by the theory, in which the mentioned collective excitations do not interact with each other, but instead the corresponding one-particle Hamiltonian has an unusual form-it is defined on the Brillouin zone with the branch cut. Accross this branch cut the Hamiltonian is not continuous. It becomes continuous if the Brillouin zone is enlarged and contains several folds connected at the branch cut.
{The topological invariant responsible for the AFQHE has the same expression as Eq. (\ref{calN}) \cite{Volovik0,Volovik2003} for the AQHE. The corresponding integral over all folds of the multi-valued Hamiltonian (i.e. over several copies of the Brillouin zone) is equal to an integer. Correspondingly, an integral over the single Brillouin zone with the branch cut is equal to this integer divided by the number of folds. This is how in our approach the Hall conductivity appears to be equal to the fractional number \zz{times} Klitzing constant.}

\red{It is worth mentioning that the unusual quantum mechanics appears, when one considers time evolution of wave function under the action of the Hamiltonian with discontinuity along a curve in momentum space (See Appendix II). Namely, the wave function, which is originally smooth as a function of momenta acquires the discontinuity along the branch cut of the Hamiltonian as a result of the evolution in time. Along this cut the wave function is not smooth. However, if one would translate the wave function of this type to the language of coordinate space, the discontinuity disappears. Such a wave function remains single-valued in coordinate space. As a price for this it does not behave at spatial infinity as a wave function of a bound state. However, quasiparticles in crystals do not have to exist in bound states. The above observation only means that the very special type of elongated, nonlocalized states of quasiparticles are present in the considered phenomenological model of fractional Chern insulator.}

\red{We would like to notice that there exists the particular well - known textbook case, when interactions lead to the appearance of the fermionic propagator with the branch cut in momentum space. This is the case of QED plasma. In the presence of exchange by photons at finite temperature the Green function for Dirac electrons becomes a nontrivial  function of absolute value $p$ of momentum and frequency $\omega$. This is a function containing logarithm of $p-\omega$ and $p+\omega$, which is multi-valued. There is a certain indirect analogy here with our case of the Hamiltonian, which is a multivalued function of momenta components. Based on this analogy one may suppose, that under certain circumstances interactions may cause the direct appearance of the effective multi-valued Hamiltonian. }

A possible interpretation of the many-fold Hamiltonian is that it describes the  degeneracy of the ground state of the system. The quasiparticles described by this Hamiltonian live within the Brillouin zone with the branch cut. Those are the excitations above one of the degenerate ground states of the system. Crossing the branch cut we drop to the fold of the Brillouin zone, which corresponds to the quasiparticles around another representative of the ground states set. Correspondingly, the $n$-fold one-particle Hamiltonian may model the excitations above the $n$-fold degenerate ground states. Having in mind all mentioned above we come to the following pattern. {In the conventional systems} without interactions the electrons have single-valued one-particle Hamiltonians, which give rise to the integer value of the topological invariant of Eq.  (\ref{calN}). {Due to the presence of interactions or for other reasons} the fermionic excitations {in fractional Chern insulators} above the ground state may be described, at least, qualitatively, by the Brillouin zone with a branch cut. The one-particle Hamiltonian of these excitations is multi-valued. Across the branch cut there is the entrance to the world of the excitations above another copy of the ground state. Since only one of the degenerate ground states is realized in reality, the one-particle excitations only above one of the ground state copies is realized. We assume that the mentioned excitations carry the electric charge of electrons. Then the direct calculation of the   Hall current gives us the standard expression of Eq. (\ref{calN}), in which the value of $N_3$ is fractional because the one-particle Hamiltonian is multi-valued.
{This pattern may appear, for example, due to the interactions between the ordinary Bloch electrons with single-valued Hamiltonians. The interactions then make out of the Bloch electrons the more complicated collective excitations, which already have the multi-valued Hamiltonians.
\add{In addition, in Appendix IV we consider the typical pattern of such Hamiltonians in coordinate space. It appears that the Hamiltonian should necessarily be highly non - local, which indicates that it might appear as a result of long - range Coulomb interactions. The more simple single - valued Hamiltonian of Appendix III also gives transitions not only from the nearest neighbors. However, its form remains local (there are no transitions between infinitely distant lattice points).}

 Alternatively, the lattice Bloch electrons themselves may acquire the multi-valued Hamiltonians in certain materials without any relation to Coulomb interactions, or exchange by quanta of lattice excitations. In both cases the particular mechanism is unknown to us, and we only conjecture its result - the appearance of the branch cut in the Brillouin zone of the model describing the fermionic excitations in FCI. }

\add{We would like to mention here the composite fermion model that has been developed originally for the description of FQHE in the presence of magnetic field.
For the case of the AFQHE the analogue of the composite fermion model has  been developed in \cite{Fradkin2}. In fractional Chern insulators the (AF)QHE appears without external magnetic field. Its role is played by the band topology. Therefore, the emergent rather than real electromagnetic field is related to magnetic fluxes that form composite fermions being attached to electrons. There may be certain links between the pattern proposed in \cite{Fradkin2} and our quasiparticles described by multi – particle Hamiltonian. However, the investigation of such links is far out of the scope of our paper. }

\add{By construction the quasiparticles described by our multi – valued Hamiltonians carry electric charge of electrons, and are described by the Grassmann – valued field. The latter means that we deal with the fermions. Thus the fractional statistics as well as the fractional charge do not appear in this pattern. However, it is possible to suppose that our non – interacting model with multi – valued Hamiltonian should be supplemented with extra interactions between the quasiparticles. We may also assume that these interactions do not renormalize Hall conductivity just like weak interactions that do not renormalize the IQHE on the level of perturbation theory. However, we do not exclude that as a result of interactions the nature of the true excitations may be changed so that these true excitations will have both fractional statistics and fractional charges.}

\red{As a possible extension of our present construction we would like to mention the case of the spin Hall effect. As well as the QHE conductivity, the spin Hall conductivity may be expressed through the Green functions in the way similar to that of the Hall conductivity. The only difference is the presence of a matrix that takes into account the spin degrees of freedom. If this matrix commutes or anti-commutes with the Green function, the corresponding expression is an integer-valued topological invariant. Again, the Hamiltonian with a branch cut would give rise to fractional value of this quantity. }

\zz{It is worth mentioning, that the topological invariance is lost, strictly speaking, in the case of multi-valued Green functions. For integer AQHE the topological invariant $\cal N$ of Eq. (\ref{calN}) is robust to smooth variations of the Green function. If the Green function is multi - valued this property is lost: arbitrary variations of $\cal G$ may lead to a change in the value of $N_3$. If, however, the integral in Eq. (\ref{calN}) is extended to all folds of the multi-valued Green function (giving an integer number), this property will be back. Thus the breakdown of topological invariance is due to the possibility that integrals in Eq. (\ref{calN}) over different folds of the multi - valued Hamiltonian give different values. Arbitrary variations of $\cal G$ may cause this. Let us consider the systems with the Hamiltonian of the form of Eq. (\ref{HknH1}). Then Eq. (\ref{N3Hknsgn}) prompts when the topological invariance is restored for the proposed expression of Hall conductivity. This occurs if we restrict our consideration to those variations of the Hamiltonian, for which the branching points (the endpoints of the branch cuts) remain coinciding with the positions of zeros of function $h_1(p) + i h_2(p)$. It is also necessary that the type of the singularities at the branching point remains the same (i.e. if we turn around the branching point, function $h_1(p) + i h_2(p)$ acquires the same phase as without modification). $N_3$ remains robust to such variations (for the detailed consideration see Appendix I).}

{Notice, that in our consideration we still did not take into account both interactions between the quasiparticles, and disorder. It is natural to suppose, that the only modification of an expression for the Hall conductivity is (as in the case of integer AQHE) that the non-interacting two-point Green function is to be substituted by the complete interacting one. The absence of corrections to the AFQHE conductivity containing the multi-leg Green functions may be, possibly, proved extending the approach of \cite{ZZ2019} to the systems with the multi-valued Hamiltonians. Even more challenging is the possible extension of our scheme to the non-homogeneous systems. Here we expect that taking into account disorder in FCI and the consideration of the FQHE in the presence of magnetic field may be achieved using the same methodology related somehow to the one of \cite{ZW2019,FZ2019_2,FZ2019,ZZ2019_2,FSWZZ2019}. However, this extension of our research remains out of the scope of the present paper. }








\section*{Appendix I: Robustness of fractional Hall conductivity against perturbations}
\zz{In this appendix we establish conditions under which variations do not lead to change of Hall conductivity. First of all let us consider} the fractional winding number defined in Eq. (\ref{N'1}). The variation of $ N'_1$ is
\begin{align}
	\delta N'_1[z',\bar{z}']&=\frac{1}{4\pi i}\Big(\int_{t=t_0}^{t=t_0+T} (z'(t)+\delta z'(t))^{-1}d (z'(t)+\delta z'(t))\nonumber\\
	&(\bar{z}'(t)+\delta \bar{z}'(t))^{-1}d (\bar{z}'(t)+\delta \bar{z}'(t))\Big)\nonumber\\
	&-\frac{1}{4\pi i}\Big(\int_{t=t_0}^{t=t_0+T} (z'(t))^{-1}d z'(t)+(\bar{z}'(t))^{-1}d \bar{z}'(t) \Big)
\end{align}
for infinitesimal $\delta z'(t)$ and $\delta \bar{z}'(t)$. Since
\be
	&&(z'(t)+\delta z'(t))^{-1}\approx z'(t)^{-1}(1-z'(t)^{-1}\delta z'(t))\nonumber\\
	&&(\bar{z}'(t)+\delta \bar{z}'(t))^{-1}\approx \bar{z}'(t)^{-1}(1-\bar{z}'(t)^{-1}\delta \bar{z}'(t))\,,
\ee		
\begin{align}
	\delta N'_1[z',\bar{z}']&\approx \frac{1}{4\pi i}\Big(\int_{t=t_0}^{t=t_0+T} z'(t)^{-1}d\delta z'(t)-z'(t)^{-2}\delta z'(t)d z'(t)\nonumber\\
	&+ \bar{z}'(t)^{-1}d\delta \bar{z}'(t)-\bar{z}'(t)^{-2}\delta \bar{z}'(t)d \bar{z}'(t)\Big)\nonumber\\
	&=\frac{1}{4\pi}\Big(\int_{t=t_0}^{t=t_0+T} d(\delta z'(t)z'(t)^{-1}+\delta \bar{z}'(t)\bar{z}'(t)^{-1}\Big)\nonumber\\
	&=\frac{1}{4\pi}\Big(\delta z'(t)z'(t)^{-1}+\delta \bar{z}'(t)\bar{z}'(t)^{-1}\Big)\Big|_{t=t_0}^{t=t_0+T}\,.
\end{align}
We know that
\be
	 \left\{
  \begin{array}{lr}
    z'(t+T)=z'(t)e^{ \frac{2\pi n_1i}{n}} &\\
    \bar{z}'(t+T)=\bar{z}'(t)e^{-\frac{ 2\pi n_1i}{n}} &
  \end{array}
\right.\,
\ee
\zz{Therefore, in general $\delta N'_1$ may be nonzero.}
However if we require the same periodicity for $\delta z'(t)$ and $\delta \bar{z}'(t)$, that is
\be	 \left\{
  \begin{array}{lr}
    \delta z(t+T)=\delta z(t)e^{ \frac{2\pi n_1i}{n}} &\\
    \delta\bar{z}(t+T)=\delta\bar{z}(t)e^{-\frac{2\pi n_1i}{n}} &
  \end{array}
\right.,
\ee
\zz{then} $\delta N'_1=0$ for any $t=t_0$. \zz{The infinitesimal transformations of this type do not change topology of maps} \be
	 \left\{
  \begin{array}{lr}
    z'(t)=z^{\frac{1}{n}}(t) &\\
    \bar{z}'(t)=\bar{z}^{\frac{1}{n}}(t) &
  \end{array}
\right. .
\ee
Next, we study the \zz{value of $N_3$ for} a multi-valued Hamiltonian defined by Eq. (\ref{N3Hkn}) and its infinitesimal deformation.
\begin{eqnarray}
	&&\zzz{\delta N_3[H_{k/n}]}\nonumber\\
	&=&\frac{i}{8\pi}\int_{BZ}  d \Big( \frac{h_3+\delta h_3}{|h_{\delta}|} d \ln (\frac{h_1+\delta h_1+i(h_2+\delta h_2)}{h_1+\delta h_1-i(h_2+\delta h_2)})\Big)\nonumber\\
	&&-\frac{i}{8\pi}\int_{BZ}  d \Big( \frac{h_3}{|h|} d \ln (\frac{h_1+ih_2}{h_1-ih_2})\Big)\nonumber\\
\end{eqnarray}
\begin{eqnarray}
	&=&\frac{i}{8\pi}\int_{BZ}  d \Big( \frac{h_3+\delta h_3}{|h_{\delta}|} (d \ln (\frac{h_1+\delta h_1+i(h_2+\delta h_2)}{h_1+\delta h_1-i(h_2+\delta h_2)})\nonumber\\
	&&-d \ln (\frac{h_1+ih_2}{h_1-ih_2})\Big)\nonumber\\
	&&+\frac{i}{8\pi}\int_{BZ}  d \Big( ( \frac{h_3+\delta h_3}{|h_{\delta}|}- \frac{h_3}{|h|}) d \ln (\frac{h_1+ih_2}{h_1-ih_2})\Big)\,.\nonumber\\
\end{eqnarray}
The second term can be integrated \zz{using procedure of} Sec. (\ref{SectWL}) and becomes
\begin{align}
	\frac{k}{2n}\sum_{p^a}((\text{sgn}(h'_3+\delta h'_3)-\text{sgn}(h'_3))\text{sgn}(\det (\frac{\partial \tilde{h}'_i}{\partial p_j})|_{p_j=p^a_j}))\,.
\end{align}
For infinitesimal $\delta h_3$ this is zero. The first term becomes
\be
	\frac{i}{8\pi}\int_{BZ}  d \Big( \frac{h_3+\delta h_3}{|h_{\delta}|} d( (\delta h_1+i\delta h_2) (h_1+ih_2)^{-1}\nonumber\\
	+ (\delta h_1-i\delta h_2) (h_1-ih_2)^{-1})\,.
\ee
For this to vanish, we need to require similar conditions as in the case of $\delta z'(t)$ and $\delta \bar{z}'(t)$. Around every singularity,
if \be
	h_1(\theta_0+2\pi)\pm i h_2(\theta_0+2\pi)=( h_1(\theta_0)\pm ih_2(\theta_0))e^{\pm \frac{2\pi }{n}i}\nonumber\\
\ee
we require
\begin{eqnarray}
&&	\delta h_1(\theta_0+2\pi)\pm i\delta h_2(\theta_0+2\pi)\nonumber\\&&=(\delta h_1(\theta_0)\pm i\delta h_2(\theta_0))e^{\pm \frac{2\pi }{n}i}\, \label{Acond}
\end{eqnarray}
to make the second term \zz{vanishing. Here $\theta$ is an angle corresponding to turning around the given branching point. $\theta_0$ is an initial value of $\theta$, while $\theta_0+2\pi$ is its value after the complete circle around the branching point. This way the topology of $h_1\pm i h_2$ is unchanged, and $N_3[H_{k/n}]$ remains robust to the variations of the Hamiltonian of this type.}

\zzz{It is worth mentioning that the variations of Hamiltonian considered above not necessarily correspond to fixed positions of the branching points in momentum space. The variations may result in moving of those points, and the condition of Eq. (\ref{Acond}) corresponds to turning around the new (moved) branching point. We would also like to notice that the variation satisfying this condition is not smooth everywhere. Being the multi - valued function it has to be discontinuous along the branch cut.}

\section*{Appendix II: Time evolution of the branched Hamiltonian}
In this appendix, let us show that the time evolution of any state can be carried by the two eigenstates within one Riemann sheet of the Hamiltonian. Let's consider Hamiltonian Eq. (\ref{H13}). The discussion of the general Hamiltonian Eq. (\ref{HknH1}) is completely parallel.
The eigenvalues of  Eq. (\ref{H13})
\be
	H_{1/3}&=&\left[\begin{array}{cc}m & (p_1-ip_2)^{1/3} \\(p_1+ip_2)^{1/3} & -m\end{array}\right]\nonumber ,
\ee
are
\be
	E_{\pm}=\pm \sqrt{(p_1^2+p_2^2)^{1/3}+m^2}\,,
\ee
whose  two orthonormal eigenvectors are
\be\label{ev}
	| \pm (p_1,p_2) \rangle =\frac{1}{\sqrt{2E^2_{\pm}-2mE_{\pm}}}\left(\begin{array}{c} (p_1-ip_2)^{1/3}\\m-E_{\pm}\end{array}\right)\,.
\ee
Once we determine the Riemann sheet of $ (p_1-ip_2)^{1/3}$, the eigenvectors are also determined. There are three possibilities for Eq. (\ref{ev}), namely
\be
	&0\le \theta(p_1,p_2)<\frac{2\pi}{3}\nonumber\\
	\text{ or } &\frac{2\pi}{3}\le \theta(p_1,p_2)< \frac{4\pi}{3}\nonumber\\
	\text{ or } &\frac{4\pi}{3}\le \theta(p_1,p_2)< \frac{6\pi}{3}\,,
\ee
where
\be
	(p_1^2+p_2^2)^{1/6}e^{-i\theta(p_1,p_2)}=(p_1-ip_2)^{1/3}\,.
\ee
Let's choose in the following as a convention the eigenvectors in which $0\le \theta(p_1,p_2)<\frac{2\pi}{3}$ are satisfied.
Any state $$| \alpha(p_1, p_2) \rangle =\left(\begin{array}{c}\xi_{\alpha}(p_1,p_2) \\ \eta_{\alpha}(p_1,p_2)\end{array}\right)$$ can be expanded by these two eigenvectors:
\be
	| \alpha(p_1, p_2) \rangle=c_{+} | + (p_1,p_2) \rangle +c_{-} | - (p_1,p_2) \rangle\,.
\ee
Taking inner product of $| \alpha(p_1, p_2) \rangle$ with each of the eigenvectors, we get the coefficients $c_+$ and $c_-$ respectively, with the branch also determined:
\be
	\langle + | \alpha \rangle = c_+=\frac{(p_1+ip_2)^{1/3} \xi_{\alpha}+(m-E_+)\eta_{\alpha}}{\sqrt{2E^2_{+}-2mE_{+}}}\,, \\
	\langle - | \alpha \rangle = c_-=\frac{(p_1+ip_2)^{1/3} \xi_{\alpha}+(m-E_-)\eta_{\alpha}}{\sqrt{2E^2_{-}-2mE_{-}}}\,.
\ee
The time evolution in Schrodinger picture is clear. We
consider time evolution operator
\be
	\mathcal{U}(t)=\exp (-iH_{1/3}t)\,
\ee
and because the Hamiltonian is Hermitian, the unitarity of time evolution is always satisfied. For any state we can write
\be
&&	| \alpha (t; p_1,p_2) \rangle = \mathcal{U}(t) | \alpha (p_1,p_2) \rangle \nonumber
	\\
	&&=\exp (-iE_{+}t)c_+|+ (p_1,p_2) \rangle\nonumber\\&&+\exp (-iE_{-}t)c_-|- (p_1,p_2) \rangle \nonumber
	\\
	&&=\exp (-iE_{+}t)\Big(c_+|+ (p_1,p_2) \rangle\nonumber\\&&+\exp (i(E_{+}-E_{-})t)c_-|- (p_1,p_2) \rangle\Big)\,.
\ee
This means that the time evolutions for different Riemann sheets slightly differ from each other. \red{Moreover, if the original wave function is smooth in momentum space, at finite $t$ it acquires the discontinuity along the branch cut of function $(p_1+ip_2)^{1/3}$. One can easily check that the real space lattice Hamiltonian (corresponding to the given multi-valued Hamiltonian in momentum space) is essentially nonlocal, but single-valued. It is given by Fourier transformation of the function with discontinuity along a curve in Brillouin zone. Such a Fourier transformation itself does not have discontinuities in coordinate space. But it is highly non-local. }

\section*{Appendix III. Coordinate representation of single - valued Hamiltonian.}

Let us consider the toy model with the following Hamiltonian:
\be \label{Hp}
&&\mathcal{H}(p_1,p_2)=\nonumber\\&&=\left[\begin{array}{cc} M+\sum_{i=1}^{2}(1-\cos p_i) & (\sin p_1-i\sin p_2)^m \\(\sin p_1+i\sin p_2)^m & -M-\sum_{i=1}^{2}(1-\cos p_i)\end{array}\right]\nonumber\\
&&=\left[\begin{array}{cc}  \mathcal{H}_{11}&  \mathcal{H}_{12} \\ \mathcal{H}_{21} &  \mathcal{H}_{22}\end{array}\right]\,.
\ee
The field Hamiltonian in second-quantized form is given by:
\be \label{HBZ}
H=\int \frac{d^2 \bs{p}}{(2\pi)^2} ~\psi^{\dagger}(\bs{p})\mathcal{H}(p_1,p_2)\psi(\bs{p})\,,
\ee
where
\be
\psi (\bs{p})=\left(\begin{array}{c} \psi_1 (\bs{p})\\ \psi_2(\bs{p}) \end{array}\right)\,.
\ee
Let us represent transform this Hamiltonian to coordinate space.

Each component of the above Hamiltonian is a sum of polynomials of $e^{\pm ip_i}$ representing lattice translations or hoppings between adjacent sites. The off-diagonal components represent the  sum of the terms corresponding to the jumps between the nearest neighbors and also between the sites with the distance between them up to m lattice spacings.

We represent the Hamiltonian as follows:
\be \label{Hx}
H&=&\sum_{\bs{x}}\psi_{\bs{x}}^{\dagger} \mathcal{H}(\hat{p}_1,\hat{p}_2)\psi_{\bs{x}}\nonumber
\\
&=&\sum_{\bs{x}}\Big(\psi_{1,\bs{x}}^{\dagger} (M+\sum_{i=1}^{2}(1-\cos\hat{p}_i) )\psi_{\bs{1,x}}\nonumber\\&& +\psi_{1,\bs{x}}^{\dagger} (\sin \hat{p}_1-i\sin \hat{p}_2)^m \psi_{\bs{2,x}} \nonumber
\\&&+\psi_{2,\bs{x}}^{\dagger}(\sin \hat{p}_1+i\sin \hat{p}_2)^m \psi_{\bs{1,x}}\nonumber\\&&+ \psi_{2,\bs{x}}^{\dagger}(-M-\sum_{i=1}^{2}(1-\cos \hat{p}_i))\psi_{\bs{2,x}}\Big)\nonumber\\
&=&H_{11}+H_{12}+H_{21}+H_{22}\,,
\ee
where
\be
\psi_{\bs{x}}=\left(\begin{array}{c} \psi_{1,\bs{x}}\\ \psi_{2,\bs{x}}\end{array}\right)\,.
\ee
It is easy to see that $e^{\pm i\hat{p}_i}$ represents the nearest neighbor hopping:
\begin{align}\label{hp1}
	(e^{i\hat{p}_1})^n \psi_{\bs{x}}&=(e^{i\hat{p}_1})^n \int \frac{d^2 \bs{p}}{(2\pi)^2}~e^{i\bs{p}\cdot\bs{x}} \psi (\bs{p})\nonumber\\
	&= \int \frac{d^2 \bs{p}}{(2\pi)^2}~e^{inp_1}e^{i\bs{p}\cdot\bs{x}} \psi (\bs{p})\nonumber\\
	&=\psi_{x_1+n,x_2}\,,
\end{align}
and similarly
\begin{align}\label{hp2}
	(e^{i\hat{p}_2})^n \psi_{\bs{x}}=\psi_{x_1,x_2+n}\,.
\end{align}
Moreover we can rewrite each component of the Hamiltonian as a sum of the powers of $e^{\pm ip_i}$:
\be\label{H1122}
\mathcal{H}_{11}(p_1,p_2)&&=m+2-\frac{1}{2}(e^{ip_1}+e^{-ip_1}+e^{ip_2}+e^{-ip_2})\nonumber\\&&=-\mathcal{H}_{22}(p_1,p_2)\,,
\ee
and
\be\label{H12}
&&\mathcal{H}_{12}(p_1,p_2)
=(\sin p_1-i\sin p_2)^m\nonumber\\
&&=\sum_{n=0}^m(-i)^n \binom{m}{n}(\sin p_1)^{m-n}(\sin p_2)^n\nonumber\\
&&=\sum_{n=0}^m(-i)^n \binom{m}{n}\sum_{j=0}^{m-n}(\frac{1}{2i})^{m-n}(-1)^j\binom{m-n}{j}e^{ip_1(m-n-2j)}\nonumber\\
&&\sum_{l=0}^{n}(\frac{1}{2i})^{n}(-1)^l\binom{n}{l}e^{ip_1(n-2l)}\nonumber\\
&&=(\frac{1}{2i})^{m}\sum_{n=0}^m(-i)^n \binom{m}{n}\sum_{j=0}^{m-n}\sum_{l=0}^{n}\nonumber\\&&(-1)^{j+l}\binom{m-n}{j}\binom{n}{l}e^{ip_1(m-n-2j)+ip_2(n-2l)}\,,\\
\label{H21}&&	\mathcal{H}_{21}(p_1,p_2)=(\sin p_1+i\sin p_2)^m\nonumber\\
&=&(\frac{1}{2i})^{m}\sum_{n=0}^m i^n \binom{m}{n}\sum_{j=0}^{m-n}\sum_{l=0}^{n}(-1)^{j+l}\nonumber\\&&\binom{m-n}{j}\binom{n}{l}e^{ip_1(m-n-2j)+ip_2(n-2l)}\,.
\ee
Substituting Eq. (\ref{hp1}, \ref{hp2}, \ref{H1122}, \ref{H12}, \ref{H21}) to Eq. (\ref{Hx}), we obtain the explicit  Hamiltonian in coordinate space:
\begin{align}
&	H_{11}=\sum_{\bs{x}}\psi^{\dagger}_{1,x_1,x_2}((M+2)\psi_{1,x_1,x_2}\nonumber\\&-\frac{1}{2}(\psi_{1,x_1+1,x_2}+\psi_{1,x_1-1,x_2}+\psi_{1,x_1,x_2+1}+\psi_{1,x_1,x_2-1}))\,,\nonumber\\
&	H_{12}=\sum_{\bs{x}}\psi^{\dagger}_{1,x_1,x_2}((\frac{1}{2i})^{m}\sum_{n=0}^m(-i)^n \binom{m}{n}\sum_{j=0}^{m-n}\sum_{l=0}^{n}(-1)^{j+l}\nonumber\\&\binom{m-n}{j}\binom{n}{l}\psi_{2,x_1+m-n-2j, x_2+n-2l}\,,\nonumber\\
&	H_{21}=\sum_{\bs{x}}\psi^{\dagger}_{1,x_1,x_2}((\frac{1}{2i})^{m}\sum_{n=0}^m i^n \binom{m}{n}\sum_{j=0}^{m-n}\sum_{l=0}^{n}(-1)^{j+l}\nonumber\\&\binom{m-n}{j}\binom{n}{l}\psi_{2,x_1+m-n-2j, x_2+n-2l}\,,\nonumber\\
&	H_{22}=\sum_{\bs{x}}\psi^{\dagger}_{2,x_1,x_2}(-(M+2)\psi_{2,x_1,x_2}\nonumber\\&+\frac{1}{2}(\psi_{2,x_1+1,x_2}+\psi_{2,x_1-1,x_2}+\psi_{2,x_1,x_2+1}+\psi_{2,x_1,x_2-1}))\,.
\end{align}
One can see that the complicated function in momentum space generates hopping parameters depending on distance in coordinate space.

\section*{Appendix IV. Coordinate representation of multi - valued Hamiltonian}
Here we consider the coordinate realization of multi - valued Hamiltonian. We take as an example
\be \label{Hf}
&&\mathcal{H}_{k/m}(p_1,p_2)\nonumber\\&&=\left[\begin{array}{cc} M+\sum_{i=1}^{2}(1-\cos p_i) & (\sin p_1-i\sin p_2)^{k/m} \\(\sin p_1+i\sin p_2)^{k/m} & -M-\sum_{i=1}^{2}(1-\cos p_i)\end{array}\right]\nonumber\\
&=&\left[\begin{array}{cc}  \mathcal{H}_{11}&  \mathcal{H}^{k/m}_{12} \\ \mathcal{H}^{k/m}_{21} &  \mathcal{H}^{}_{22}\end{array}\right]\,.
\ee
The diagonal elements are the same as in Appendix III. However, the off-diagonal elements are different.
\be\label{Hf12}
\mathcal{H}^{k/m}_{12}(p_1,p_2)
&=&(\sin p_1-i\sin p_2)^{k/m}
\ee
There are two ways to expand this term.
If
\be
\Big| \frac{\sin p_2}{\sin p_1} \Big|<1
\ee
then
\be
&&\mathcal{H}^{k/m}_{12}(p_1,p_2)=\sum_{n=0}^{\infty}(-i)^n \binom{\frac{k}{m}}{n}(\sin p_1)^{\frac{k}{m}-n}(\sin p_2)^n\nonumber\\
&&=\sum_{n=0}^\infty(-i)^n \binom{\frac{k}{m}}{n}\sum_{j=0}^{\infty}(\frac{1}{2i})^{\frac{k}{m}-n}(-1)^j\binom{\frac{k}{m}-n}{j}e^{ip_1(\frac{k}{m}-n-2j)}\nonumber\\
&&\sum_{l=0}^{n}(\frac{1}{2i})^{n}(-1)^l\binom{n}{l}e^{ip_2(n-2l)}\nonumber\\
&&=(\frac{1}{2i})^{\frac{k}{m}}\sum_{n=0}^\infty(-i)^n \binom{\frac{k}{m}}{n}\sum_{j=0}^\infty\sum_{l=0}^{n}(-1)^{j+l}\nonumber\\&&\binom{\frac{k}{m}-n}{j}\binom{n}{l}e^{ip_1(\frac{k}{m}-n-2j)+ip_2(n-2l)}\,,\label{exp1}
\ee
but if
\be
\Big| \frac{\sin p_2}{\sin p_1} \Big|>1
\ee
then the following expansion is to be used
\be
&&\mathcal{H}^{k/m}_{12}(p_1,p_2)=\sum_{n=0}^{\infty} \binom{\frac{k}{m}}{n}(\sin p_1)^n(i\sin p_2)^{\frac{k}{m}-n}\nonumber\\
&&=\sum_{n=0}^\infty i^{\frac{k}{m}-n} \binom{\frac{k}{m}}{n}\sum_{l=0}^{n}(\frac{1}{2i})^{n}(-1)^l\binom{n}{l}e^{ip_1(n-2l)} \nonumber\\
&&\sum_{j=0}^{\infty}(\frac{1}{2i})^{\frac{k}{m}-n}(-1)^j\binom{\frac{k}{m}-n}{j}e^{ip_2(\frac{k}{m}-n-2j)}\nonumber\\
&&=(\frac{1}{2i})^{\frac{k}{m}}\sum_{n=0}^\infty i^{\frac{k}{m}-n} \binom{\frac{k}{m}}{n}\sum_{j=0}^\infty\sum_{l=0}^{n}(-1)^{j+l}\nonumber\\&&
\binom{\frac{k}{m}-n}{j}\binom{n}{l}e^{ip_1(n-2l)+ip_2(\frac{k}{m}-n-2j)}\,,\label{exp2}
\ee
One can see that the two expansions are in general different if $k/m$ is a rational number. This reflects the presence of the branch cut in the Brillouin zone. In Fig. \ref{fig001} we represent the Brillouin zone of this model. Inside it there are four squares, where $|{\rm sin} p_1| < |{\rm sin} p_2|$ and four squares, where
$|{\rm sin} p_1| > |{\rm sin} p_2|$. Those squares are separated by the lines $p_1 = \pm p_2, p_1 = \pm \pi \pm p_2$.

\begin{figure}[h!]   \centering     \includegraphics[width=0.6 \columnwidth]{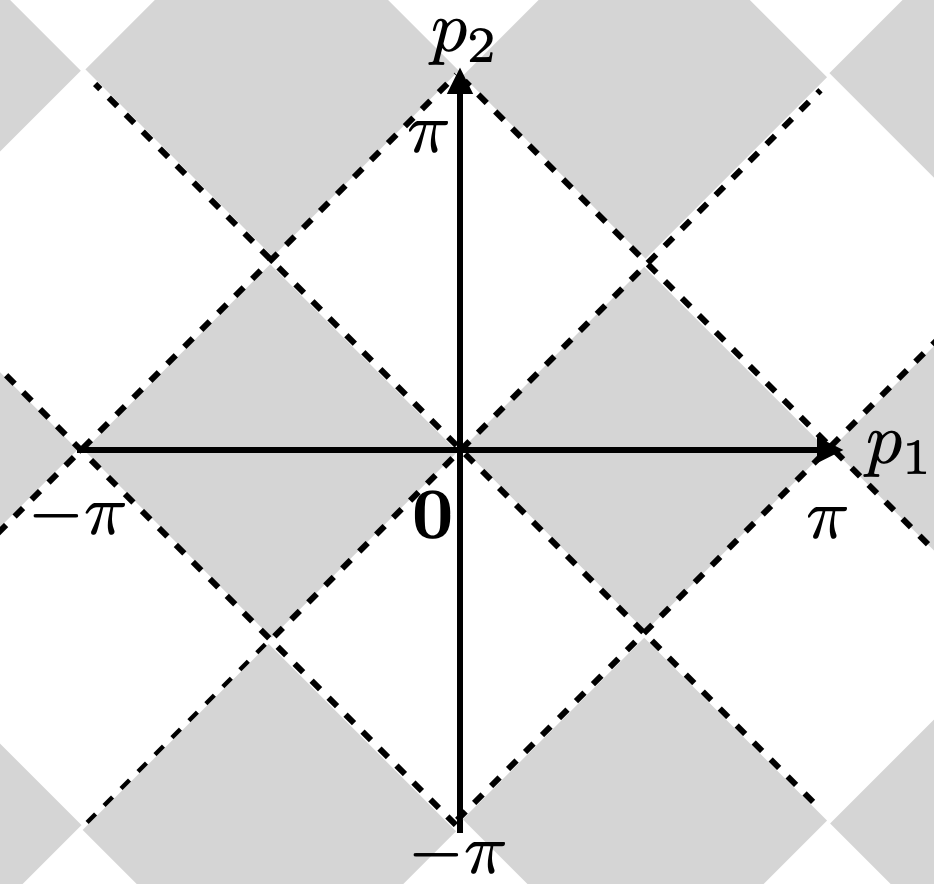}     \caption{ \label{fig008}{In the figure the regions of different expansions for the model of Appendix IV are represented using different colors. }.  } \end{figure}

If there would be only one expansion of Eq. (\ref{exp2}) everywhere in the Brillouin zone, then the formal expression for the coordinate space Hamiltonian looks as
\be
&&H^{k/m}_{12}=\sum_{\bs{x}}(\frac{1}{2i})^{\frac{k}{m}}\sum_{n=0}^\infty(-i)^n \binom{\frac{k}{m}}{n}\sum_{j=0}^\infty\sum_{l=0}^{n}\nonumber\\&&(-1)^{j+l}\binom{\frac{k}{m}-n}{j}\binom{n}{l}\nonumber\\
&&\psi^{\dagger}_{1,x_1,x_2}\psi_{2,x_1+\frac{k}{m}-n-2j, x_2+n-2l}\,,\nonumber\\
\ee
However, since Eq. (\ref{exp2}) works only in half of the Brillouin zone, calculating the Fourier transform (an integral over momenta) we should take into account also the expansion of Eq. (\ref{exp1}). The whole integration region should be separated into the mentioned above 8 squares, in four of them we use Eq. (\ref{exp2}), in another four we use Eq. (\ref{exp1}) (see Fig. \ref{fig008}). The resulting expression is rather complicated, and we do not present it here. However, the main feature of this expression is its essential non - locality. Namely, unlike the case of the model of Appendix III, here the hopping parameters are present for the transitions between arbitrarily distant points of the lattice.

We conclude that only the lattice model with non - locality (with the couplings between distant points) may lead to the multi - valued Hamiltonians.  Another important feature of such systems originates from the discontinuity of momentum space Hamiltonian along the branch cut. The derivative of the Hamiltonian with respect to momentum has delta - functional singularity along the branch cut.


\vskip2pc


\end{document}